\documentclass[11pt,a4paper]{emulateapj}
\usepackage{epsfig}
\usepackage{amsmath}
\usepackage{natbib}
\usepackage{graphics}
\usepackage{color}

\usepackage{hyperref}
\hypersetup{
 bookmarksnumbered=true,
 pdftitle = {What could we learn from a sharply falling positron fraction?}, 
 pdfauthor = {Timur Delahaye, Kumiko Kotera, Joseph Silk}, 
 pdfsubject = {positron fraction}, 
 pdfkeywords = {positron fraction, AMS-02, dark matter, cosmic rays},
 colorlinks=true,
 linkcolor=blue,
 citecolor=blue,
 urlcolor=blue,
}

\begin{document}

\title{What could we learn from a sharply falling positron fraction?}

\author{Timur Delahaye\altaffilmark{1,2,3}, Kumiko Kotera\altaffilmark{1,4}}
\author{Joseph Silk\altaffilmark{1,5,6}}

 \altaffiltext{1}{
 Institut d'Astrophysique de Paris
UMR7095 -- CNRS, Universit\'e Pierre \& Marie Curie,
98 bis boulevard Arago
F-75014 Paris, France.
}%
 
\altaffiltext{2}{
LAPTH, Universit\'{e} de Savoie, CNRS; 9 chemin de Bellevue, BP110, F-74941 Annecy-le-Vieux Cedex, France 
}%

\altaffiltext{3}{
Oskar Klein Centre for Cosmoparticle Physics, Department of Physics, Stockholm University, SE-10691 Stockholm, Sweden
}%

\altaffiltext{4}{ 
Department of Physics and Astronomy, University College London,
Gower Street, London WC1E 6BT, United Kingdom
}%

\altaffiltext{5}{ 
The Johns Hopkins University, Department of Physics and Astronomy, 3400 N. Charles Street, Baltimore, Maryland 21218, USA
}%

\altaffiltext{6}{  
Beecroft Institute of Particle Astrophysics and Cosmology, Department of Physics, University of Oxford, Denys Wilkinson Building, 1 Keble Road, Oxford OX1 3RH, UK}%

\begin{abstract}
Recent results from the AMS-02 data have confirmed that the cosmic ray positron fraction increases with energy between 10 and 200~GeV. This quantity should not exceed 50\%, and  it is hence expected that it will either converge towards 50\% or fall. We study the possibility that future data may show the positron fraction dropping down abruptly to the level expected with only secondary production,  and forecast the implications of such a feature in term of possible injection mechanisms that include both dark matter and pulsars. { Were a sharp steepening to be found, rather surprisingly, we conclude that pulsar models would do at least as well as dark matter scenarios in terms of accounting for any spectral cut-off.}
\end{abstract}

\maketitle

\section{Introduction}

The positron fraction, that is, the flux of cosmic-ray positrons divided by the flux of electrons and positrons, has attracted much interest since the publication of the results of the PAMELA satellite \citep{Adriani2009,Adriani2013}. 
PAMELA has indeed reported an anomalous rise in the positron fraction with energy, between 10 and 200~GeV. These measurements have been confirmed recently by AMS-02~\citep{Aguilar2013}.  The intriguing question is what may happen next? The positron fraction  must either saturate or decline. In the latter case, how abrupt a decline might we expect? The naive expectation is that a dark matter self-annihilation interpretation, bounded by the particle rest mass, should inevitably generate a sharper cut-off than any astrophysical model.

Antiparticles are rare among cosmic rays, and can be produced as {\it secondary} particles by cosmic ray nuclei while they propagate and interact in the interstellar medium. The sharp increase observed in the positron fraction is however barely compatible with the most simple models of secondary production. Various alternatives have been proposed, such as a modification of the propagation model~\citep{Katz2009,Blum2013}, or primary positron production scenarios, with pulsars~(e.g., \citealp{Grasso09,Hooper09,Delahaye2010,Blasi11,Linden13}) or dark matter annihilation~(e.g., \citealp{Delahaye2008,Arkani-Hamed09,Cholis09,Cirelli09}) as sources. The current data and the uncertainties inherent in the source models do not yet enable us to rule out these scenarios. It is however likely that improved sensitivities at higher energies and a thorough measurement of the shape of the spectrum above $\sim 200\,$GeV will be able to constrain the models. This question has been studied in earlier work (see for instance \citealp{Ioka2010,Kawanaka2010,Pato2010,Mauro2014}); here we want to test more specifically the possibility of a sharp drop of the positron fraction. An original aspect of our work is  to also convolve our results with the cosmic-ray production parameter space for pulsars allowed by theory.

The AMS-02 data presents a hint of flattening in the positron fraction above 250 GeV. Such a feature is expected, as the positron fraction should not exceed 0.5, and hence it should either converge towards 0.5 or start decreasing. We investigate in this paper the following question: what constraints could we put on dark matter annihilation and primary pulsar scenarios if the next AMS-02 data release were to show a sharply dropping positron fraction? A sharp drop could be deemed natural if the positron excess originates from the annihilation of dark matter particles with a mass of several hundred GeV. However, we show in this work that such a feature would be highly constraining in terms of dark matter scenarios. 
More unexpectedly, we demonstrate that pulsar models could also lead to similar results for a narrow parameter space. Interestingly, we find that pulsars lying in this parameter space happen to be the only ones that would be astrophysically capable of contributing to the pair flux at this level. 

In this paper, we first describe our method and our assumptions, then we analyse the dark matter and pulsar scenarios respectively. Finally, we discuss our results.

\section{Method} 
\label{sec:method}

\begin{figure}[t]
\centering
\includegraphics[width=0.45\textwidth]{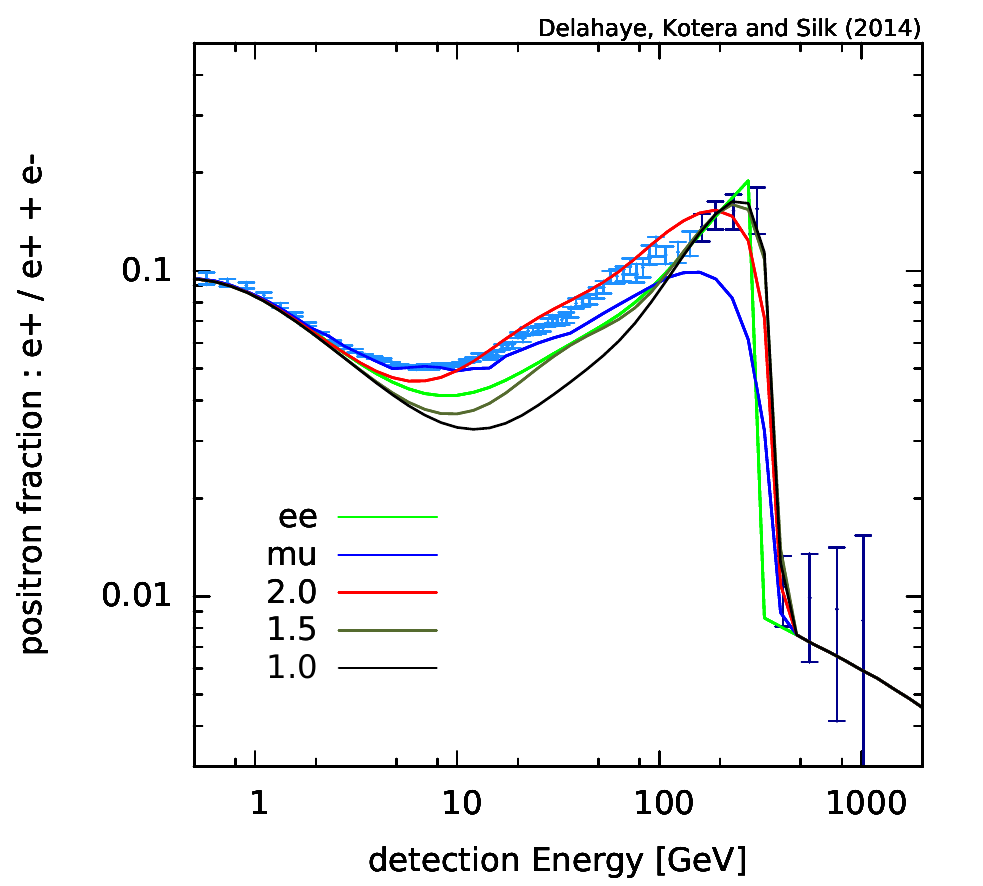}
\includegraphics[width=0.45\textwidth]{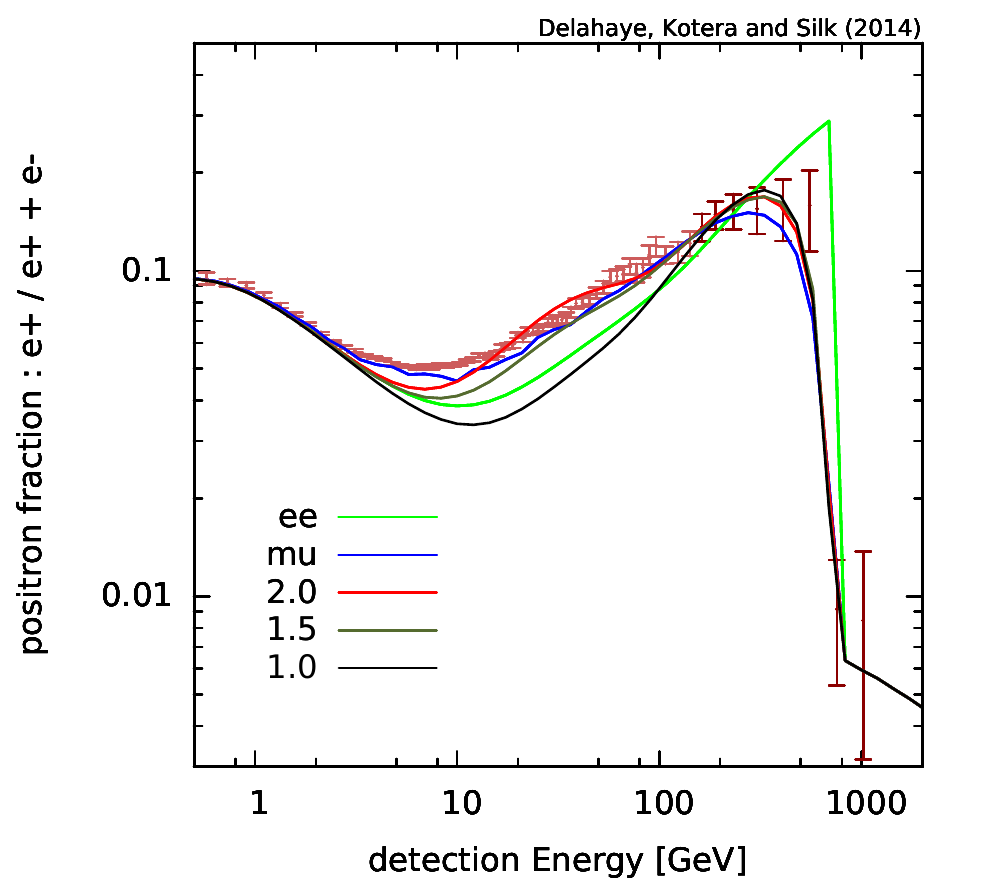}
\caption{
Best fit fluxes for the {\it max} parameter set. Upper panel for a positron drop at 350~GeV, lower panel at 600~GeV. Data up to 350~GeV is from AMS-02~\citep{Aguilar2013}, above this energy, the bins are mock data. The lines correspond respectively to dark matter annihilating into $e^+e^-$ or $\mu^+\mu^-$ or to a pulsar with injection spectrum \textbf{parameter} of 1, 1.5 or 2.
Note that for the pulsar cases, a smooth distribution of far away pulsars, with the same injection spectrum (but a lower cut-off) has been added to reproduce the data at intermediate energies (10 to 150 GeV).
}
\label{fig:fluxes}
%
\vskip -0.25cm %
%
\end{figure}

In order to mimic a sharp drop in the positron fraction measurements, we generate two sets of mock data by extrapolating the AMS-02 data points at higher energies. We assume that the flux keeps rising up to 350~GeV and 600 GeV respectively, and then drops to the level expected for a flux produced purely by secondary cosmic rays (see Fig.~\ref{fig:fluxes}). The relative error bars are assumed to be the same as the last published bin until the possible drop and to increase by 50\%\footnote{\textbf{This choice of 50\% is arbitrary. Indeed it is difficult to estimate what will be the error bars of future data as the AMS-02 collaboration has not yet published estimates of their systematics. Also, the statistical errors will depend on the electron background and on the choice of the collaboration for the energy binning. We have tested the effects of being more conservative and not increasing the relative errors does not change our results.}} at each energy bin after the drop, where the statistics will necessarily be lower for some years. We deliberately adopt such a sharp drop to test the sharpest situation possible that is commonly considered as a dark matter smoking gun.

To fit this mock data, three components can be considered: i) a standard underlying secondary flux (produced by interactions of primary cosmic rays in the interstellar medium, and inferred from observed cosmic-ray fluxes), ii) far away pulsars likely to contribute to the electron and positron fluxes between a couple of GeV and $\sim$150~GeV \citep{Delahaye2010}, iii) in addition, one can add the contribution of another primary electron and positron flux, coming either from a single nearby pulsar or from the Galactic dark matter halo. 

We compute the flux and distribution of primary and secondary cosmic rays in the Galaxy following the commonly used two-zone diffusion model, where the stars and Interstellar Medium (ISM) lie in an infinitely thin disk embedded in a large diffusion halo of chaotic magnetic field. Once in the diffusion zone, cosmic rays suffer diffusion, energy losses (mainly inverse Compton and synchrotron; note that Klein-Nishina effects are here taken into account), spallation on the ISM, convection and reacceleration. The latter two effects have not been taken into account here, as they impact only low energy electrons and  we are interested here in energies above $\sim$200~GeV. The various parameters that quantify these phenomena are not known from first principles and must be constrained by data, such as the boron-to-carbon ratio, that is not sensitive to source modeling. As \cite{Maurin2001} has shown, the parameter space compatible with the data is very large and translates into an equally large uncertainty on the expected positron and electron fluxes that should be sized correctly \citep{Delahaye2008,Delahaye2009}. In order to account for this spread, we will discuss our scenarios within three sets of representative parameters labelled {\it min}, {\it med} and {\it max} in \cite{Donato2004} (Table~\ref{tab:param}).

\begin{table}[t]
  \caption{Propagation parameter sets} \label{tab:param}
  \begin{center}
    \leavevmode
\begin{tabular}{lcccc}
\hline \hline
 &min & med & max \\ 
\hline 
$L$ [kpc] & 1 & 4 & 15 \\ 
$K_0$ [kpc$^2$/Myr] & 0.0016 & 0.0112 & 0.0765 \\ 
$\delta$ & 0.85 & 0.70 & 0.46 \\ 
\hline

    \end{tabular}
  \end{center}
  {Propagation parameter sets for the three representative cases discussed in \cite{Donato2001}. $L$ is the diffusion halo size, $K_0$ the diffusion coefficient at 1\,GeV, and $\delta$ the diffusion coefficient power law index: $K(E)=K_0(E/1\,{\rm GeV})^\delta$.}

\end{table}

For each of the three propagation parameter sets, we have calculated the underlying secondary positron flux (channel i) as in Ref.~\cite{Delahaye2009}. In this study, we have chosen to use the primary proton and $\alpha$ fluxes given by~\cite{Donato2009} and the production cross-sections of~\cite{Kamae2006,Kamae2007}; other choices are possible but such a parameter scan is not within  the scope of this paper. The electron injection flux is set to follow a power-law that gives a good overall fit to the PAMELA electron data, as well as to the AMS-02 positron fraction data below 5~GeV. Note that electrons are mainly primary cosmic-rays and are hence less straightforward to model than positrons.

For illustrative purposes, in the pulsar case, we have overlaid a second component of distant pulsars (channel ii), following the spatial distribution of~\cite{Lorimer2004}. We use the the same injection power-law as in channel i, but with a lower cut-off energy, set to minimize the $\chi^2$ between 10 and 150~GeV. This has no impact on the results as this quantity is added after the parameters of the local pulsars are set. A similar exercise could be done for the dark matter scenario by adding flatter annihilation channels like $W^+W^-$.

At the highest energies, we overlay the contribution of an additional primary flux (channel iii) due to either a dark matter halo, or a single nearby pulsar. For the Dark Matter halo scenario, we have restricted our discussion to the two cases where dark matter fully annihilates into electron and muon pairs. Indeed the other annihilation channels do not lead to a sharp electron spectrum. The annihilation cross-section and the mass of the dark matter particle are left free. We assumed a NFW profile for the halo, but the choice of the profile does not affect much the result. As shown by~\cite{Delahaye2008}, positrons suffer energy losses in the Galaxy and the flux measured at the Earth cannot be affected by the exact dark matter distribution at the Galactic Centre.

In the pulsar scenario, we assume an injection at the single nearby source of the form
$Q(E) = Q_0\,E^{-\sigma} \exp(-E/E_{\rm c})$
with $E_{\rm c}$ the cut-off energy and $Q_0$ defined via the total energy ${\cal E}_{\rm tot}=\int EQ(E)\,{\rm d}E$ integrated between $E_{\rm min}=0.1\,$GeV and $E_{\rm max}=10\,E_{\rm c}$. For the spectral index $\sigma$, we have considered 3 benchmark values (see section~\ref{sec:discussion}): $1.0$, $1.5$ and $2.0$. The maximum energy $E_{\rm c}$, the total injected energy ${\cal E}_{\rm tot}$, as well as the time of injection (pulsar age) and the distance between the source and the Earth are left as free parameters.

Considering that between a couple of GeV and $\sim$150~GeV, far away pulsars could contribute (channel ii), we have required our additional primary source to give a good fit above 150~GeV only (\textit{i.e.} the four highest energy bins of the AMS-02 data and the four mock data points generated as explained previously). The source parameters are then scanned over until the $\chi^2$ is minimised. For each case, we also have computed the maximal value of the anisotropy in order to check that it always remained lower than the maximal value set by AMS-02: 0.036. The best fit results are shown in Tables~\ref{tab:low} and \ref{tab:high}.

Concerning the anisotropy, one might note that the choice of the AMS-02 collaboration to give the anisotropy of the positron ratio $\Delta$ (the flux of positron divided by the flux of negative electrons only) instead of the anisotropy in the positron flux (noted $A$ in the next section) is surprising, as it is \textbf{less} constraining, compared to individual anisotropies. Indeed, whatever the source of the positron excess it should produce electrons in the same quantity and hence the anisotropy of the positron ratio is expected to be small, even when the positron flux is very anisotropic. 

\section{Results}

\begin{table}[t]
  \caption{Best $\chi^2$ for pulsar scenarios, drop energy 350~GeV} \label{tab:low}
  \begin{center}
    \leavevmode
\begin{tabular}{lrrr}
\hline\hline
$\chi^2$\quad$[\Delta/10^{-4}]$ & min & med & max \\ 
\hline 

$\sigma=1.0$ & 0.6 [1.7] & 0.4 [0.3]  & 0.7 [0.2] \\  
$\sigma=1.5$ & 1.6 [3.3] & 1.0 [3.3]  & 1.1 [3.3] \\ 
$\sigma=2.0$ & 2.6 [4.7] & 2.8 [4.4]  & 6.6 [3.5] \\ 
$\mu^+\mu^-$ & 20.3 & 27.0 & 61.5 \\
$e^+e^-$ & 5.3 & 1.6 & 13.7\\
\hline
\end{tabular} 
  \end{center}
  {Best $\chi^2$ for pulsar scenarios with $\sigma=[1,1.5,2]$ (first three lines), and for dark matter scenarios, for the three propagation parameter sets defined in Table~\ref{tab:param} for a drop energy of 350~GeV and in brackets, the corresponding highest anisotropy signal $\Delta$ in units of $10^{-4}$. \textbf{The number of degrees of freedom is 8.}}

\end{table}

\begin{table}[t]
  \caption{Best $\chi^2$ for pulsar scenarios, drop energy 600~GeV. } \label{tab:high}
  \begin{center}
    \leavevmode
\begin{tabular}{lrrr}
\hline\hline
$\chi^2$\quad$[\Delta/10^{-3}]$ & min & med & max \\ 
\hline 
$\sigma=1.0$ & 2.5 [1.1] & 2.8 [0.6] & 3.8 [0.5]\\ 
$\sigma=1.5$ & 3.2 [5.1] & 2.7 [4.7] & 2.2 [3.0]\\ 
$\sigma=2.0$ & 4.4 [8.4] & 4.1 [8.0] & 3.3 [8.7]\\ 
$\mu^+\mu^-$ & 4.8 & 2.7 & 5.7 \\
$e^+e^-$ & 27.6 & 17.8 & 13.7\\
\hline
\end{tabular} 

  \end{center}
  Same as Table~\ref{tab:low}, but for a drop energy of 600~GeV. \textbf{Again, the number of degrees of freedom is 8.}\\

\end{table}

Tables~\ref{tab:low} and \ref{tab:high} present the best results of our $\chi^2$ analysis for a sharp drop of the positron fraction at 350 and 600~GeV respectively. The tables display the values of the best-fit $\chi^2$ (together with the anisotropy signal $\Delta$) for each benchmark case. The corresponding fluxes for the {\it max} propagation case are shown in Fig.~\ref{fig:fluxes}.

\subsection{Dark Matter}

Not very surprisingly, in the case of a dark matter annihilating solely into $e^+ + e^-$ it is possible to obtain a sharp drop of the positron fraction for the {\it min} and {\it med} propagation parameters. In the {\it max} case the  $\chi^2$ is less good though. Note that this annihilation channel is quite sharp on both sides of the fall and does not reproduce very well a fraction that would flatten before dropping (\textit{high} case).

In the case where dark matter would annihilate into $\mu^+ + \mu^-$, however, the fraction is smoother before the drop which gives a better result in our \textit{high} case.

Note however that in all cases, the annihilation cross-sections (or boost factors) required to fit the data are very high. This is already known for quite some time and raises a large number of issues concerning consistency of such a results with other observations such as anti-protons~\citep{Donato2009}, $\gamma$-rays~\citep{Cirelli:2009dv}, synchrotron emission~\citep{Linden:2011au} etc.

\subsection{Pulsars}

\begin{figure*}[t]
\centering
\includegraphics[width=.48\textwidth]{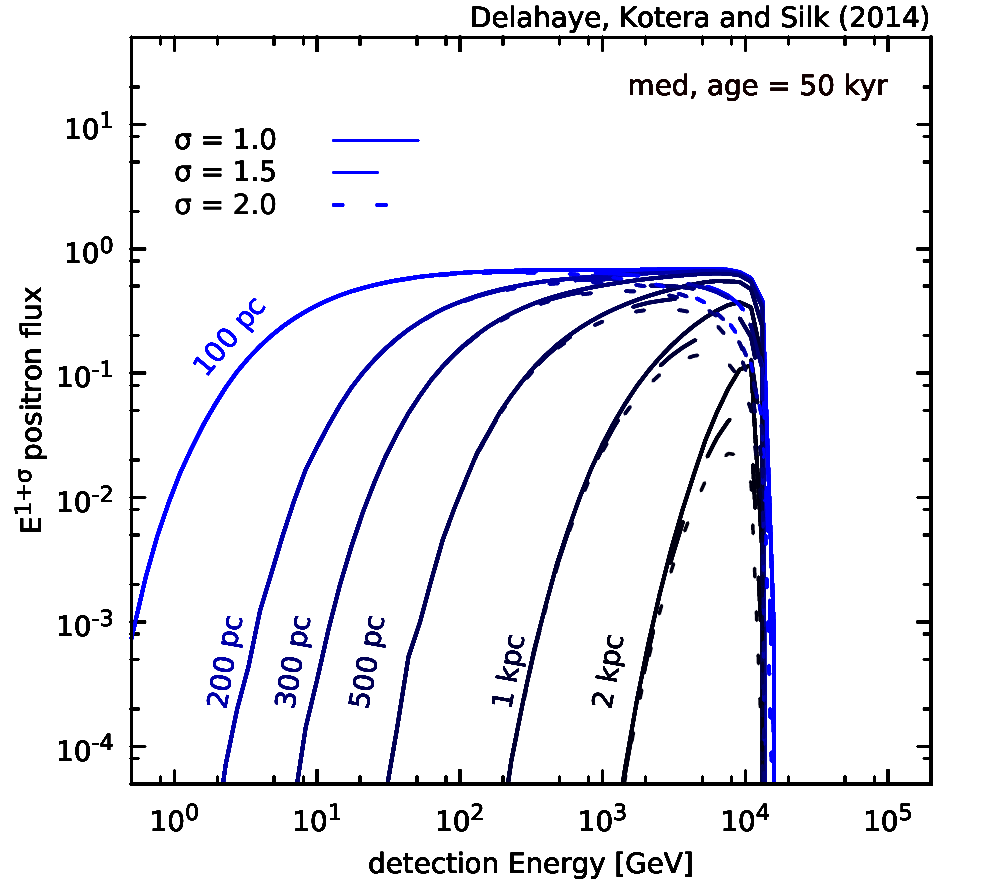}
\includegraphics[width=.48\textwidth]{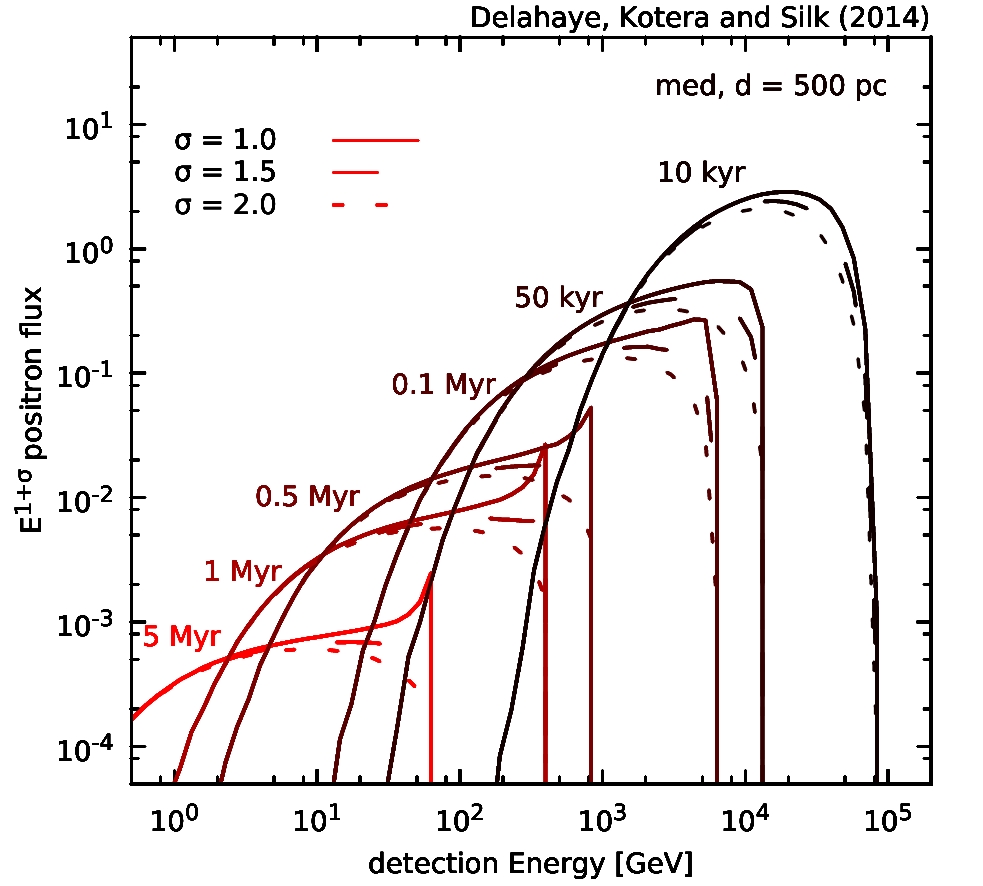}
\caption{
Impact of the distance (left panel) and of the age (right panel) of a pulsar on the positron flux received at the Earth. In order to show only the effects of the propagation, the injection energy cut-off has been set to the very high value of 100~TeV for all the cases. Continuous, dashed and dotted lines correspond to an injection respectively of $\sigma=$1, 1.5 and 2. 
The fluxes displayed here are corrected by a factor $E^{1+\sigma}$ to ease the comparison.
It clearly appears that distance has little impact on the shape of the flux at high energies. One should also note that the flux coming from old pulsars drops more sharply, whatever the distance. \\}
\label{fig:pulsars}
%
\vskip -0.25cm %
%
\end{figure*}

\begin{figure*}[ht!]
\centering
\includegraphics[width=.98\textwidth]
{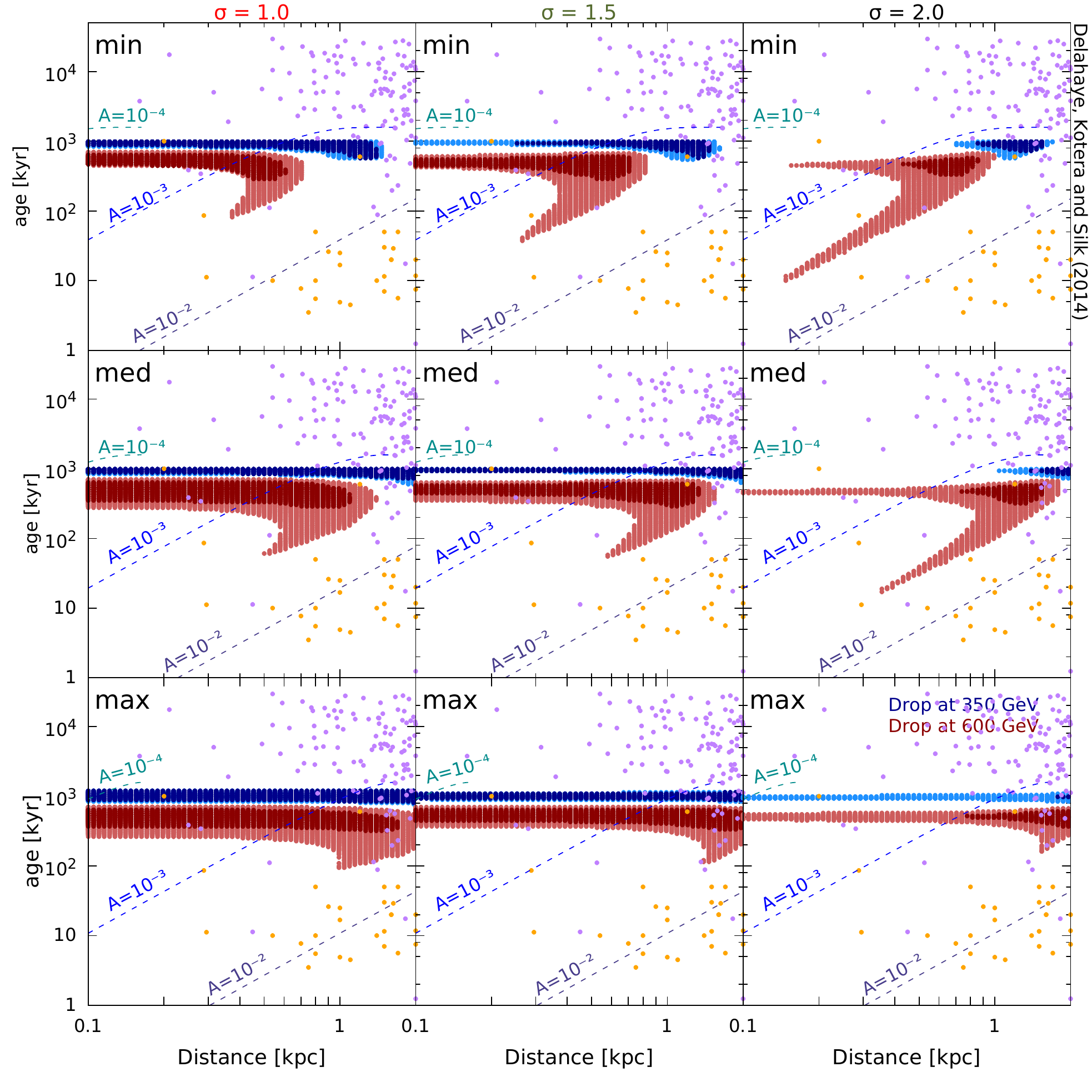}
\caption{
{\it min} {\it med} {\it max} (top to bottom), injection spectral index $\sigma=1.0, 1.5$ and 2.0 (left to right) for a sharp drop at 350~GeV (in blue) or at 600 GeV (in red). Dark areas correspond to pulsars leading to $\chi^2<8$ (the number of degrees of freedom) whereas  the shaded area represent the $2-\sigma$ contours around the best fits given in Tables~\ref{tab:low} and~\ref{tab:high}. The blue dashed lines give an estimate of the anisotropies in the positron flux {$A$ (not in the positron fraction)} induced by a unique pulsar sitting at a given distance and time. This is not a full calculation but an analytical estimate where the secondary background is neglected and the pulsar sits in the direction of the Galactic Center. The energy cut-off at injection $E_{\rm c}$ was left as a free parameter and is displayed in Fig.~\ref{fig:carre}.
The purple and orange dots correspond to the existing cosmic ray sources one can find in the ATNF~\citep{Manchester2005} and Green~\citep{Green2009} catalogues. \\}
\label{fig:novuple}
%
\vskip -0.25cm %
%
\end{figure*}

\begin{figure}[ht!]
\centering
\includegraphics[width=.5\textwidth]{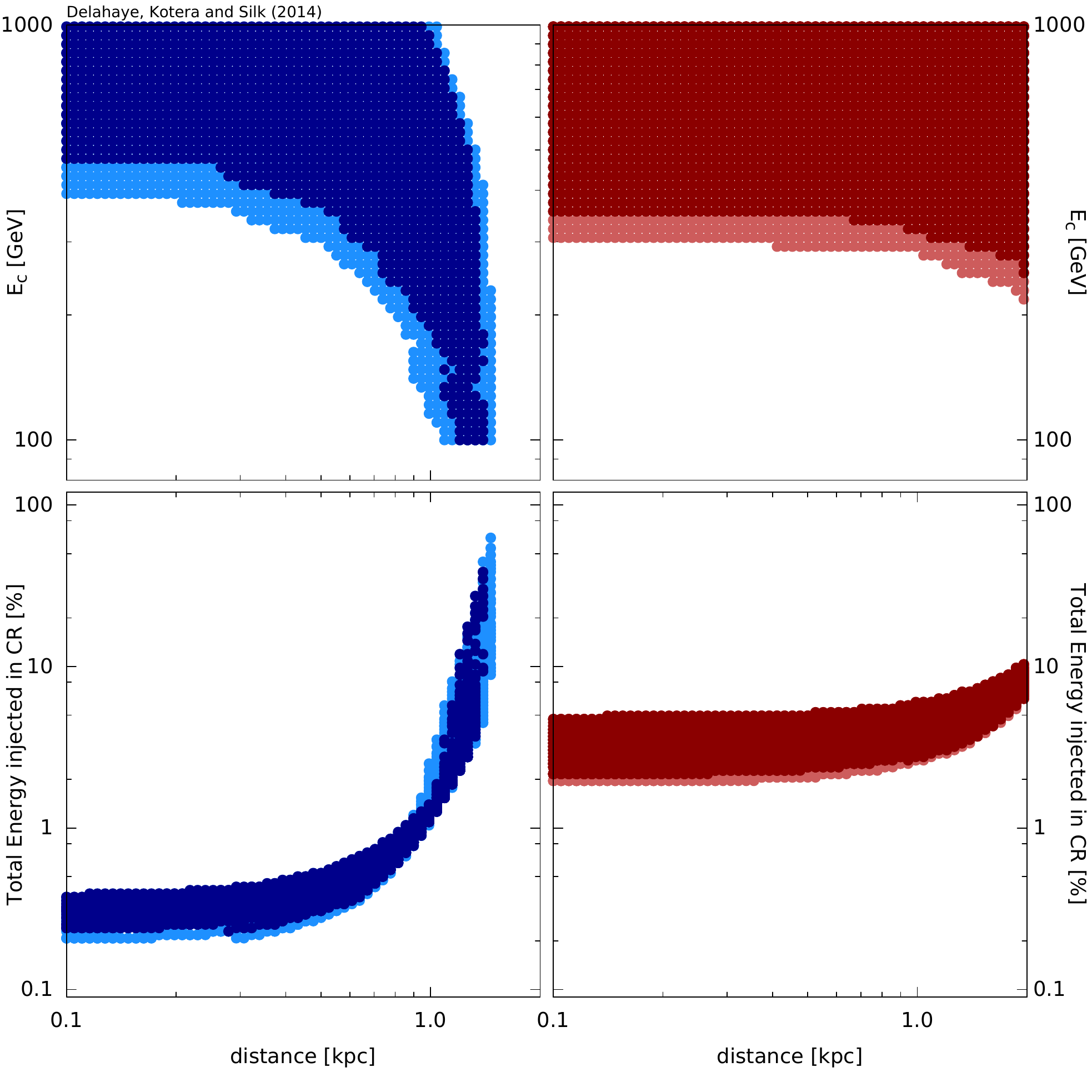}
\caption{
Energy cut-off at injection (top) and total energy going to cosmic-ray (bottom) for an injection spectrum $\sigma=1$. The left-hand panels correspond to the {\it min} case and a drop at 350~GeV, whereas the right-hand panels are for a drop at 600~GeV and the {\it max} propagation parameters.\\
}
\label{fig:carre}
%
\vskip -0.25cm %
%
\end{figure}

Our best $\chi^2$ values indicate that some scenarios exist, where single pulsars can lead to a good fit to the data. 
In order to better assess the allowed parameter space, we perform a broader parameter scan on the pulsar distance, age, and cut-off energy. Our results are presented in Figs.~\ref{fig:novuple} and~\ref{fig:carre}, where the contours represent the regions where the $\chi^2$ is lesser than the number of degrees of freedom of the fit (dark colors) and 2-$\sigma$ away from the best fit value (light shades). The bottom plots in Fig.~\ref{fig:carre} display the corresponding energy injected into cosmic rays ${\cal E}_{\rm tot}$, as a fraction of a typical supernova explosion energy (10$^{51}$ erg). The relevance of this energy budget is discussed in section~\ref{sec:discussion}. 

The shape of the spectra produced by single pulsars is an intricate combination of injection parameters and propagation effects. The influence of the various quantities are discussed in detail in \cite{Delahaye2010}. Figure~\ref{fig:pulsars} recalls the effects of the pulsar distance and age on the observed spectral slope and high-energy cut-off. 

The position of the observed cut-off in energy is set by one of the two following effects. Either it is set solely by the age of the pulsar, given an initial maximum energy $E_{\rm c}$; the narrow horizontal bands of fixed pulsar age in Fig.~\ref{fig:novuple} correspond to this effect. The higher drop-energy ($600\,$GeV) case is naturally better fit by younger pulsars than the 350\,GeV case. Or, in other cases, the observed cut-off is due to the maximum energy at which cosmic rays are accelerated, $E_{\rm c}$ this gives the diagonal departure from the horizontal of Fig.~\ref{fig:novuple}.  One can also note from Fig.~\ref{fig:pulsars} that older pulsars lead to a sharper cut-off, whatever the distance.  

The other parameters (distance and spectral index) govern the spread of the spectrum and its steepness, down to low energies. Shorter distances and harder spectra lead to more peaked spectra, as required for our fits. These two parameters have opposing effects on the  normalization: the flux amplitude decreases with the source distance and increases for harder spectral indices. This explains why, for softer spectral indices, nearby pulsars are excluded, as they are not able to provide enough energy to account for the observed flux. For the {\it min} propagation case, large distances are excluded for this same reason, as indicated in Fig.~\ref{fig:carre}, where the fraction of injected energy saturates at 100\% for large distances. 

The dashed lines in Fig.~\ref{fig:novuple} provide an estimate of the anisotropy (in the case where the pulsar would be the only source and would sit in the Galactic plane). This is the positron flux anisotropy, as opposed to  the positron ratio reported by AMS-02. It shows that a sharp positron fraction does not necessarily imply a high anisotropy.

The diffuse cosmic-ray flux scales roughly as $\propto L/K(E)$, where $L$ is the halo size and $K(E)$ the diffusion coefficient. The {\it min} propagation case is thus intrinsically favoured energetics-wise. Additionally, this case can lead to a narrower peaked spectrum. {The anisotropy of the positron flux has however an inverse scaling $A \propto K(E)/L$, which explains why the anisotropy constraint is strongest for the {\it min} case, where $A$ is largest. Note also that here all the pulsars have been considered to be in the Galactic plane but should a pulsar be above or below the plane, the anisotropy would increase, especially in the  \textit{min} case.}\\

To summarize, two regimes appear from Fig.~\ref{fig:novuple}: a good fit to the sharp drop requires either a relatively old pulsar (horizontal branches of the scatter plots) and then the break is set by the age of the pulsar, independently from the injection cut-off and its distance, or one requires a relatively young and nearby pulsar (diagonal branch in the upper right panel). The parameter space that enables a good fit shrinks considerably as the injection index $\sigma$ increases.

\section{Discussion} 
\label{sec:discussion}

In the pulsar framework, our parameter scan favours a relatively old (a few hundred kyr old) close-by source (within $\sim 1$\,kpc), capable of supplying at least ${\cal E}_{\rm tot}\sim 10^{47-48}\,$erg into electrons and positrons, accelerated with a hard spectrum. This parameter scan was performed taking into account only propagation arguments. 
We discuss in this section how likely such a single source scenario is from an astrophysical point of view, in terms of energy budget and given the actual pulsar population. 

Pair production and acceleration in pulsars happens in several steps: electrons are initially stripped off the surface of the star by strong rotation-induced electric fields and undergo electromagnetic cascading in a yet unidentified region, which could be the polar cap \citep{Ruderman75}, the outer gap \citep{Cheng86}, or the slot gap \citep{Harding06}. The produced pairs are then channelled into the pulsar magnetosphere, and can either escape following open field lines \citep{Chi96}, or reach the pulsar wind nebula (PWN), a shocked region at the interface between the wind and the supernova ejecta, where particles can be further accelerated to high energies. 

Most pulsars are born with rotation periods $\sim 300\,$ms \citep{Lorimer2008}, which implies a rotational energy budget of $\sim 10^{46-47}\,$erg. Unless a fair fraction of the supernova ejecta energy is injected into particle kinetic energy, it is thus difficult to account for ${\cal E}_{\rm tot}\sim 10^{47-48}\,$erg required to fit the observed flux for the majority of pulsars. Pulsars that could supply this amount of energy should thus be rare sources, either because they need to spin faster, or because the conversion of the ejecta energy into particle kinetic energy has to be highly efficient. 

As long as the pulsar wind is embedded in the supernova remnant, the accelerated pairs lose energy adiabatically via expansion and radiatively via interactions with the magnetic and radiative fields. \cite{Blasi11} shows however that accelerated pairs can escape in the interstellar medium if they are liberated after the pulsar escapes the parent supernova remnant. This event typically occurs $50$\,kyr after the initial blast, as can be estimated by assuming an average birth kick velocity of the pulsar. Thus pulsars younger than this age would be naturally ruled out as contributors to the rising positron flux, as they would not have escaped the remnant yet, and accelerated particles would be trapped. 

On the other hand, older pulsars cannot contribute to the high-energy end of the spectrum either, because of propagation effects (Fig.~\ref{fig:pulsars}). Positrons produced by these sources would pile-up at intermediate energies (channel ii, mentioned in section~\ref{sec:method}).

From Fig.~\ref{fig:carre}, one can see that the bulk of the more distant pulsars $\gtrsim 1\,$kpc demand that an (unreasonably) large energy budget be channelled into cosmic rays. A typical pulsar beyond 1\,kpc can contribute at a level of $<1\%$ of the flux of a more nearby pulsar, and hundreds of sources would be needed to reach the same level of flux as one pulsar at a distance closer than 1\,kpc. Note also that most of these far away pulsars have an injection cut-off $E_{\rm c}$ that is lower than the observed cut-off. This surprising fact can happen because the injection cut-off has been set as an exponential function, and this means that some cosmic rays are also accelerated to higher energies. However this would also mean that these pulsars will become much brighter in the future when the cosmic rays with energies below the cut-off will finally reach us. This makes these source even more unlikely to explain a sharp drop of the positron fraction.

{The anisotropy of the positron flux $A$ should be stronger than that of  the positron ratio $\Delta$ on which AMS-02 set an upper limit of 0.036 and hence could be more constraining if detected. However it is not clear that even in the case of a bright source dominating the signal that it would be strong enough to impose a  strong conclusion. The energy dependence of $A$ could help as we expect the anisotropy to increase together with the flux in the case of a single pulsar dominating the signal, whereas if the dark matter halo were to dominate, the anisotropy could decrease while the flux increases. This is all the more true if the propagation parameters are close to those of \textit{min}, as this would increase the energy dependence of the anisotropy. The direction of the anisotropy could also be useful if the pulsar or pulsars responsible for the signal are not in the direction of the Galactic Center. Indeed the pulsar scenario would ultimately be satisfactory only if the brightest pulsar is actually identified and detected.}

Finally, the energy spectrum injected by a single pulsar depends on the environmental parameters of the pulsar. The toy model of unipolar induction acceleration in pulsars would lead to a hard spectral slope of index $\sigma \sim 1$ \citep{Shapiro83}. More detailed models by \cite{Kennel84a} suggest that the pair injection spectrum into the pulsar wind nebula should present a Maxwellian distribution due to the transformation of the bulk kinetic energy of the wind into thermal energy, and a non-thermal power-law tail formed by pairs accelerated at the shock. Hybrid and particle-in-cell (PIC) simulations show indeed such a behaviour (e.g., \citealp{Bennett95,Dieckmann09,Spitkovsky08}), and the latest PIC simulations indicate a relatively hard spectral slope $\sigma \sim 1.5$ \citep{Sironi11} due to acceleration by reconnection in the striped wind. 

All these arguments demonstrate that the narrow parameter space pointed out by our scan is astrophysically justified a posteriori. Because such  sources should be rare, it is consistent that not more than one of them would be currently operating. The dots in Fig.~\ref{fig:novuple} confirm indeed that existing pulsars present in the allowed parameter region are scarce. \\

\acknowledgements
{In this work, we have considered the possibility that future AMS-02 data may show a positron fraction dropping down abruptly to the level expected with only secondary production, and forecast the implications of such a feature in term of possible injection mechanisms that include both dark matter and pulsars. We have shown that dark matter scenarios would then have to face strong constraints to fit the spectral shape successfully. Pulsar models could also lead to similar results for a narrow parameter space. Interestingly, we have argued that pulsars lying in this parameter space happen to be the only ones that would be astrophysically capable of contributing to the pair flux at this level. Were a sharp steepening to be found, rather surprisingly, we conclude that pulsar models would do at least as well as dark matter scenarios in terms of accounting for any spectral cut-off.}

\section*{Acknowledgements}

This work was supported in part by ERC project 267117 (DARK MATTERS) hosted by Universit\'e Pierre et Marie Curie - Paris 6. KK acknowledges financial support from PNHE and ILP. 

\bibliography{fraction}

\end{document}